\documentclass{pasj00}

\draft

\begin{document}
\SetRunningHead{S. Kato}{Disk Oscillations and Frequency Correlations among QPOs}
\Received{2011/00/00}
\Accepted{2011/00/00}

\title{An Attempt to Describe Frequency Correlations among kHz QPOs and HBOs
by Two-Armed Nearly Vertical Oscillations}

\author{Shoji \textsc{Kato}}
\affil{2-2-2 Shikanoda-Nishi, Ikoma-shi, Nara, 630-0114}
\email{kato@gmail.com, kato@kusastro.kyoto-u.ac.jp}

%


\maketitle

\begin{abstract}

We examine whether the two-armed ($m=2$) vertical p-mode oscillations trapped in the innermost
region of magnetized accretion disks with finite disk thickness can describe kHz QPOs and HBOs in LMXBs.
First, we derive the frequency-frequency correlation of the two basic oscillations (both are fundamental
modes in the vertical direction, but one is the fundamental and the other the first overtone
in the radial direction), and compare it with the observed frequency correlation of twin kHz QPOs.
Results show that the calculated frequency correlation can well describe observed correlation with 
reasonable values of parameters.
Second, we examine whether the observed frequency correlation between kHz QPOs and HBO 
can be described by regarding HBO as the first overtone oscillation in the vertical direction
(and the fundamental in the radial direction).
The results suggest that
i) innermost parts of disks on the horizontal branch are strongly diminished in their vertical thickness
(presumably by hot coronae)
 and ii) the branch is roughly a sequence of variations of magnetic fields or disk temperature.

\end{abstract}

\section{Introduction}

In neutron-star low-mass X-ray binaries (NS LMXBs) kilo-hertz quasi-periodic oscillations (kHz QPOs)
are often observed.
In many cases they appear in pairs, and temporal changes of their frequencies are correlated.
In Z-sources of LMXBs, horizontal branch oscillations (HBOs) are also observed and their frequency
change is correlated with those of kHz QPOs (Psaltis et al. 1999).
Studies on origins of kHz QPOs and HBOs and on the cause of their correlations are of importance,
since they will bring about a deep understanding of the innermost part of disks surrounding compact
relativistic sources and of mass and spin of these sources.

There are many models of kHz QPOs and HBOs, but there is still no common consensus 
on their origins.
One of the possible origins of kHz QPOs is trapped disk-oscillations excited in the inner region of
the disks.
Among many trapped oscillation modes in disks, two-armed vertical p-mode oscillations are a good candidate
of kHz QPOs (Kato 2011b, referred hereafter as paper I).
This is because in the case where disks are threated by toroidal magnetic fields, i) these oscillations
have frequencies on the order of kHz QPOs, and
ii) can describe the correlated time variation of observed twin kHz QPOs, if the strength of the 
toroidal magnetic fields varies with time.\footnote{
In this paper we show that the causes of the correlated time variation is not necessarily time variation
of magnetic fields, but temperature variation and time variation of disk thickness due to
truncation by corona can also describe the correlation.
}

To reinforce the above model of kHz QPOs, however, some issues remain to be examined.
That is, we should study how time change of disk structure other than time change of toroidal magnetic
fields affect the correlation.
In other words, we must examine how the correlation curve calculated in paper I is changed or modified 
if time variations of disk structure due to other causes are considered.

As time changes of disk structure other than a change of toroidal magnetic fields, two possibilities are
conceivable.
One is time variation of disk temperature.
It is simply expected, since a change of mass accretion rate brings about a change of disk temperature.
Another possible time variation of disk structure is a time change of disk thickness.
The cool geometrically thin disks in NS LMXBs will be surrounded by hot corona, and the transition
hight to the corona will change by change of disk state.
Hence, it will be of importance to examine whether and how the correlation curve calculated by
paper I is modified by time change of disk temerature or transition height.
The first purpose of this paper is to examine this issue.

The second purpose of this paper is to examine whether the observed correlation between 
kHz QPOs and HBOs can also be described in the framework of 
vertical p-mode oscillations, by regarding HBOs as some higher modes of the oscillations.
The results of examination suggest that this possibility may not be unrealistic, 
if the vertcal thickness of the innermost part of disks is strongly diminished in the state where
they are on the horizontal branch (HB) .

\section{Disk Models and Parameters Describing Disks}

To describe disks and their oscillations we adopt the Newtonian formulation, 
except the effects of general relativity are taken into account in 
the radial distributions of $\Omega(r)$, $\kappa(r)$, and $\Omega_\bot(r)$,
where $\Omega(r)$ is the angular velocity of disk rotation, $\kappa(r)$ and $\Omega_\bot(r)$ are,
respectively, the epicyclic frequencies in the radial and vertical directions
[see Okazaki et al. 1987 for $\kappa(r)$, and Aliev \& Galtov 1981 and Kato 1990 for $\Omega_\bot(r)$].
$\Omega(r)$ is approximated by the angular velocity of the Keplerian rotation, $\Omega_{\rm K}(r)$, since we are
considering geometrically thin disks.
Here, $r$ is the radial coordinate of the cylindrical coordinates ($r$, $\varphi$, $z$) whose $z$-axis is 
perpendicular to the unperturbed disk plane and the origin is at the disk center.

The unperturbed disks are axisymmetric with purely toroidal magnetic fields $\mbox{\boldmath $B$}_0
(r,z)$:
\begin{equation}
      \mbox{\boldmath $B$}_0(r,z)=[0,B_0(r,z),0].
\label{2.1}
\end{equation} 
The disks are assumed to be isothermal in the vertical direction,
and $\mbox{\boldmath $B$}_0(r,z)$ is
distributed in the vertical direction in such a way that the Alfv\'{e}n speed, $c_{\rm A}$, 
is constant in the vertical direction, i.e., $(B_0^2/4\pi\rho_0)^{1/2}=$ const. in the vertical
direction, where $\rho_0(r,z)$ is the density in the unperturbed disks.
Then, the hydrostatic balance in the vertical direction gives that $\rho_0(r,z)$ and
$B_0(r,z)$ are distributed in the vertical direction as (e.g., Kato et al. 1998)
\begin{equation}
    \rho_0(r,z)=\rho_{00}(r){\rm exp}\biggr(-\frac{z^2}{2H^2}\biggr), \quad{\rm and}\quad
    B_0(r,z)=B_{00}(r){\rm exp}\biggr(-\frac{z^2}{4H^2}\biggr),
\label{2.3}
\end{equation}
where the scale height $H(r)$ is related to $c_{\rm s}$, $c_{\rm A}$, and $\Omega_\bot$ by
\begin{equation}
      H^2(r)=\frac{c_{\rm s}^2+c_{\rm A}^2/2}{\Omega_\bot^2},
\label{2.4}
\end{equation}
$c_{\rm s}(r)$ being the isothermal acoustic speed.

The disk described above is assumed to be terminated at a certain height, $z_{\rm s}$, by the presence of
a hot, low-density corona.
The transition height, $z_{\rm s}(r)$, will be determined by considering thermal and hydrostatic balances
(and mass balance).
Determination of the transition height is, however, beyond the scope of this paper, and
thus it is taken as a parameter.

The above disk model which we adopt hereafter has thus three parameters to specify disk structure,
in addition to mass, $M$, and spin parameter, $a_*$, of the central source.
They are $c_{\rm s}^2(r)$, $c_{\rm A}^2(r)$, and the height of transition, $z_{\rm s}(r)$.
Instead of these parameters which have dimensions, we introduce here three dimensionless parameters.
One is $c_{\rm A}^2/c_{\rm s}^2$, and the second is $\eta_{\rm s}\equiv z_{\rm s}/H$, and
the other one is $c_{\rm s}^2/(c_{\rm s}^2)_0$, where $(c_{\rm s}^2)_0$ is a reference value of the
square of acoustic speed and we adopt the following form, based on the standard value of  
the Shakura-Shanyaev disks.

That is, in the case of non-magnetized standard Shakura-Sunyaev disks in which the gas pressure
dominates over the radiation pressure and opacity mainly comes from the free-free processes,
we have (e.g., Kato et al. 2008)
\begin{equation}
     c_{{\rm s}}^2(r)=1.83\times 10^{16}(\alpha m)^{-1/5}{\dot m}^{2/5}(r/r_{\rm g})^{-9/10}\ {\rm cm}^2\ {\rm s}^{-2},
\label{2.2}
\end{equation}
where $\alpha$ is the conventional viscosity parameter, $r_{\rm g}$ is the Schwarzschild radius
defined by $r_{\rm g}=2GM/c^2$, $m(\equiv M/M_\odot)$\footnote{
In this section and hereafter, $m$ is often used to denote $M/M_\odot$ without confusion with the
azimuthal wavenumber $m$ of oscillations.
} 
and ${\dot m}={\dot M}/{\dot M}_{\rm crit}$, ${\dot M}_{\rm crit}$ being the critical mass-accretion
rate defined by the Eddington luminosity.
For the reference value, $(c_{\rm s}^2)_0(r)$, we adopt the value of equation (\ref{2.2}) in the case of 
$\alpha=0.1$ and $\dot{m}=0.3$.
That is, we define $(c_{\rm s}^2)_0(r)$ by
\begin{equation}
   (c_{\rm s}^2)_0(r)=1.79\times 10^{16}m^{-1/5}(r/r_{\rm g})^{-9/10}\ {\rm cm}^2\ {\rm s}^{-2}.
\label{reference-speed}
\end{equation}

In summary, we specify our disk models by three dimensionless parameters:
\begin{equation}
            \beta\equiv\frac{c_{\rm s}^2}{(c_{\rm s}^2)_0}, \quad 
            \frac{c_{\rm A}^2}{c_{\rm s}^2}, \quad {\rm and}\quad  \eta_{\rm s}\equiv \frac{z_{\rm s}}{H}.
\label{parameters}
\end{equation}
Generally speaking, these parameters can be taken to be slowly varying functions of $r$ in our analyses.
Except in sunsection 4.2, however, they are taken to be constant independent of radius, since in 
the oscillations which would be related to kHz QPOs the trapped regions are narrow
(see figures 7 and 8 by Kato 2012, see also Kato 2011a).
In subsection 4.2, radial variations of $\beta$ and $c_{\rm A}^2/c_{\rm s}^2$ are briefly considered in
relation to HBOs, since in the oscillations which would be related to HBOs the trapped regions are not
always narrow (see also figures 7 and 8 by Kato 2012).

The purpose of this paper is to examine how frequencies of some basic oscillation modes of
the two-armed vertical p-mode oscillations vary as the above parameters change 
and whether correlated frequency changes of these oscillation modes can describe the observed
frequency correlations among kHz QPOs and HBOs.

\section{Radially Trapped Two-Armed Vertical p-mode Oscillations}

Disk oscillation modes in geometrically thin non-magnetized disks are usually classified by the node
number in the vertical direction, $n$, and the frequencies in the corotating frame (see, e.g., 
Kato 2001; Kato et al. 2008).
The oscillations whose radial displacement vector, $\mbox{\boldmath $\xi$}_r(\mbox{\boldmath $r$},t)$,
has no node in the vertical direction, i.e., $n=0$, are called p-mode.
Oscillations with $n\geq 1$, except for particular ones (c-mode oscillations), are classified into
g-mode and vertical p-mode oscillations.
Those whose frequencies in the corotating frame are lower are called g-mode, while those with
higher ones are vertical p-mode.

Vertical p-mode oscillations of $n=1,2,3...$ are further divided by difference of the azimuthal wavenumber,
$m$, and the node number in the radial direction, $n_r$.\footnote{
Boundary conditions also introduce further variety of solutions.
In this paper, however, we condiser only the case where the inner bounday can be taken to be a free bounday 
and outside the trapped region oscillations are evanescent (see Kato 2011a).
}
That is, the vertical p-mode oscillations are now classified by the set of ($m$, $n$, $n_r$), where
$n$ starts from 1 as $n=1,2,3...$, while $n_r$ starts from 0 as $n_r=0,1,2,..$.

In this paper, among many vertical p-mode oscillations we focus on oscillations of two-armed ($m=2$) ones, 
since when $m=2$ there are oscillation modes that are radially trapped in the innermost region of disks with
moderate frequencies (Kato 2010).
That is, two oscillations of $n_r=0$ and $n_r=1$ (both are fundamental in the vertcal direction, i.e., $n=1$) 
have frequencies comparable with those of
kHz QPOs when the disk has moderate amount of toroidal magnetic fields (Kato 2011a).
Oscillations of $n=2$, even if they have $n_r=0$, have much lower frequencies compared with those of
$n_r=1$ and $n=1$ (and thus those of kHz QPOs).
Their frequencies can become comparable with those of HBOs.
Considering these situations, we particularly pay our attention to two-armed ($m=2$) 
vertical p-mode oscillations
with the set of ($n$, $n_r$) being (1,0), (1,1), and (2,0).
It is noted that oscillations with larger $n$ and $n_r$ are not interesting observationally, since
they will not be observed with large amplitude due to phase mixing.

Eigen-functions of the above-mentioned trapped oscillations have been examined in cases
where the disks are infinitely extended isothermal ones (Kato 2011a) and ones which are terminated 
at certain heights (Kato 2012).
In the next section, we examine whether parameter dependences of these oscillations can describe
the observed frequency correlation of twin kHz QPOs and the correlation between kHz QPOs and HBOs.

\section {Frequency Correlation and Comparison with Observations}

We consider two problems separately.
First, we examine the frequency correlation between the $n_r=0$ and $n_r=1$ oscillations both with $n=1$.
This is an extension of paper I.
Second, the correlations between the $n=1$ and $n=2$ oscillations both with $n_r=0$
are considered.

\subsection{Correlation between $n_r=0$ (with $n=1$) and $n_r=1$ (with $n=1$) Oscillations}

We take the standpoint that the fundamental ($n_r=0$) and first overtone ($n_r=1$) oscillations in the radial
direction (both fundamental in the vertical direction, i.e., $n=1$) are the upper and lower kHz QPOs, respectively.
To check this possibility, we examine how the set of frequencies of these two oscillations move on a 
frequency-frequency diagram when parameters specifying disk structure change, and compare this
correlation curve on the diagram with the observed frequency-frequency plots of the twin kHz QPOs.

This comparison has already been done in paper I in the case where the disk is non-terminated isothermal one 
($\eta_{\rm s}=\infty$) with $\beta\equiv [c_{\rm s}^2/(c_{\rm s}^2)_0]=1.0$ (figure 3 in paper I) and
$\beta=3.0$ (figure 5 in paper I) by changing the strength of toroidal magnetic fields varying from 
$c_{\rm A}^2/c_{\rm s}^2=0$ to 100.
Here, with these effects included, we do more systematical comparison 
with observations by considering cases where other disk parameters are changed.

\begin{figure}
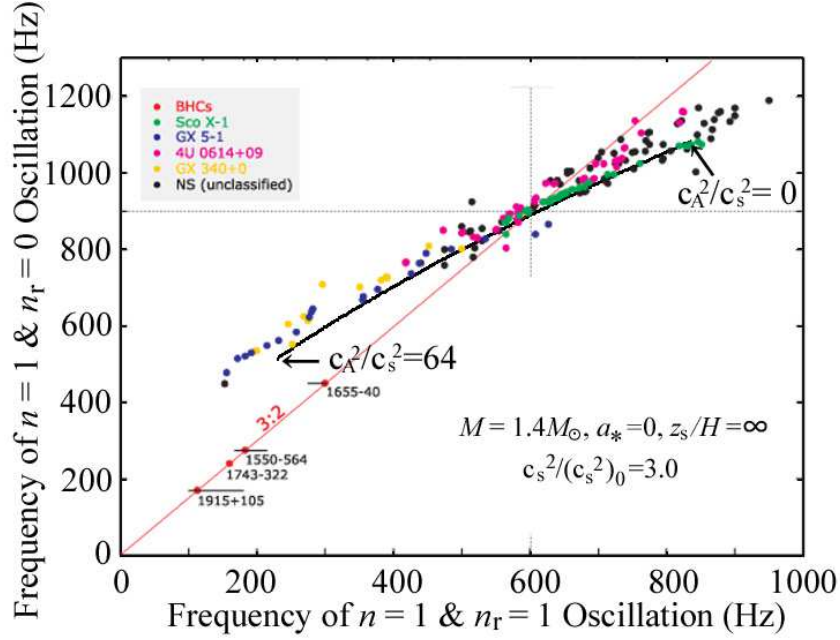

\begin{center}
    \FigureFile(120mm,120mm){figure-1.eps}
\end{center}
\caption{
Diagram comparing the calculated frequency-frequency correlation curve with the observed 
frequency-frequency plots of twin kHz QPOs of some typical NS LMXBs.
The plots of observational data are a part of the figure of Abramowicz (2005).
The straight line labelled by 3 : 2 is the line on which the frequency ratio of twin QPOs is 3 : 2.
Adopted parameters specifying disk structure are $\eta_{\rm s}=\infty$ and 
$\beta\equiv[c_{\rm s}^2/(c_{\rm s}^2)_0]=3.0$. 
The value of $c_{\rm A}^2/c_{\rm s}^2$ is changed from $c_{\rm A}^2/c_{\rm s}^2=0$ to 64.0,
and the left and right ends of the correlation curve are for $c_{\rm A}^2/c_{\rm s}^2=64.0$
and 0, respectively. (Color Online)
}
\end{figure}
\begin{figure}
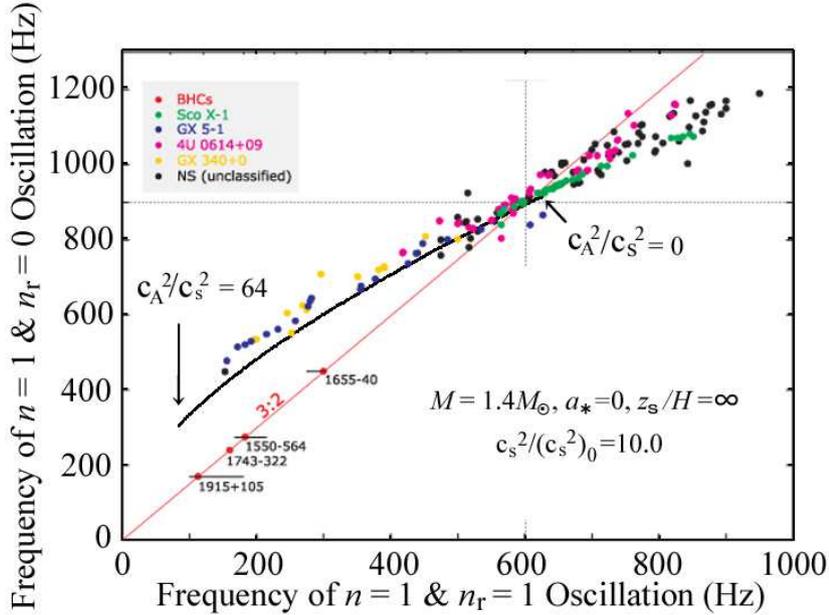

\begin{center}
    \FigureFile(120mm,120mm){figure-2.eps}
\end{center}
\caption{
The same as figure 1, except for $\beta=10.0$. (Color Online)}
\end{figure}

First, we focus on disks with $M=1.4M_\odot$ and $a_*=0$, and consider the case where
the disk is non-terminated isothermal one ($\eta_{\rm s}=\infty$).
In this case we find that the calculated correlation curve is almost universal in the sense that for a moderate
set of $\beta$ and $c_{\rm A}^2/c_{\rm s}^2$, the calculated correlation curve is almost a part of  
an unique curve.
To demonstrate this, let us first consider the case where $\beta$ is fixed at $\beta=3.0$ and 
$c_{\rm A}^2/c_{\rm s}^2$ is changed from 0 to 64, which is shown in figure 1 
(in figure 5 of paper I, $c_{\rm A}^2/c_{\rm s}^2$ is changed from 0 to 100).
When $c_{\rm A}^2/c_{\rm s}^2=0$, the point representing the set of frequencies of the $n_r=0$ (with $n=1$)
and $n_r=1$ (with $n=1$) oscillations on the frequency-frequency diagram is the right-end point
of the curve in figure 1.
In the another limit of $c_{\rm A}^2/c_{\rm s}^2=64$, the point is the left-end of the curve.
As the value of $c_{\rm A}^2/c_{\rm s}^2$ changes from 0 to 64, the point on the diagram moves
along the curve from the right-end to the left-end.
Next, let us consider the case where $\beta=10.0$ and $c_{\rm A}^2/c_{\rm s}^2$ is changed from
0 to 64, which is shown in figure 2.
Comparison of figures 1 and 2 shows that an increase of $\beta$ while keeping other parameter values
unchanged decreases frequencies of both $n_r=0$ and $n_r=1$ oscillations (see Kato 2011a).
Hence, in the case of figure 2, the correlation curve on the frequency-frequency diagram shifts towards the
left-bottom corner compared with the case of figure 1.
In the overlapping part, however, the curves in figures 1 and 2 are almost the same.

\begin{figure}
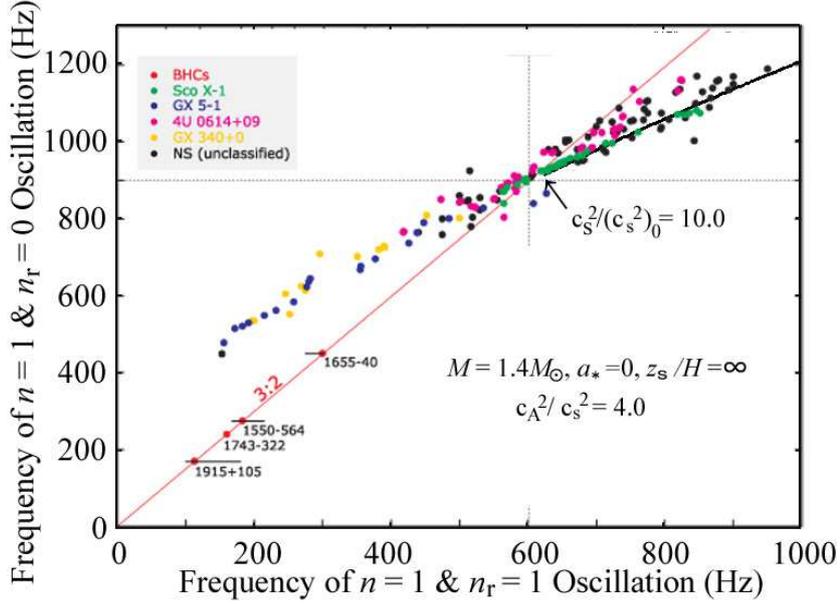

\begin{center}
    \FigureFile(120mm,120mm){figure-3.eps}
\end{center}
\caption{
The same as figures 1 and 2, except that $c_{\rm A}^2/c_{\rm s}^2$ is fixed at 
$c_{\rm A}^2/c_{\rm s}^2=4.0$ and $\beta$ is changed from $\beta=1.0$ to $\beta=10.0$.
The right-upper end-point of the correlation curve is for $\beta=1.0$, and 
the left-lower end-point of the curve is for $\beta=10.0$. (Color Online)
}
\end{figure}
\begin{figure}
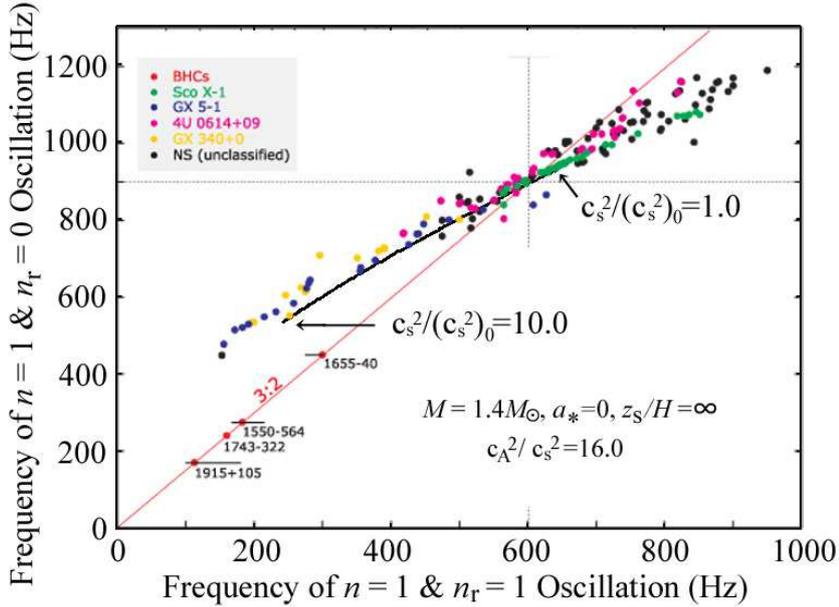

\begin{center}
    \FigureFile(120mm,120mm){figure-4.eps}
\end{center}
\caption{ 
The same as figure 3, except that $c_{\rm A}^2/c_{\rm s}^2$ is fixed at $c_{\rm A}^2/c_{\rm s}^2=16.0$.
(Color Online)
}
\end{figure}

Next, let us consider the case where $c_{\rm A}^2/c_{\rm s}^2$ is fixed and $\beta$ is changed in the 
range of $\beta=1.0\sim 10.0$, which is shown in figure 3 for $c_{\rm A}^2/c_{\rm s}^2=4.0$ and in
figure 4 for $c_{\rm A}^2/c_{\rm s}^2=16.0$.
In both figures, the left-end points of the curves are for $\beta=10.0$.
This is because if the value of $\beta$ increases while keeping other parameter values unchanged,
the frequency of both oscillations decrease (Kato 2011a).
As $\beta$ decreases, the point representing the set of the frequencies moves along the correlation curve
and reaches at the right-upper end of the curve when $\beta=1.0$.
In the case of $c_{\rm A}^2/c_{\rm s}^2=16.0$, frequencies of both oscillations are low compared
with those in the case of $c_{\rm A}^2/c_{\rm s}^2=4.0$ if other parameters are fixed.
Hence, in figure 4 the calculated correlation curve shifts to the left-lower direction compared with the
case of $c_{\rm A}^2/c_{\rm s}^2=4.0$.
However, in the region where frequencies are overlapped, the correlation curve is almost the same 
as that in figure 3.
That is, as mentioned before, the correlation curve is almost unique.
Of course, if the mass of the central source is different from $M=1.4M_\odot$ the correlation curve
on the frequency-frequency diagram is not overlapped (see below).

\begin{figure}
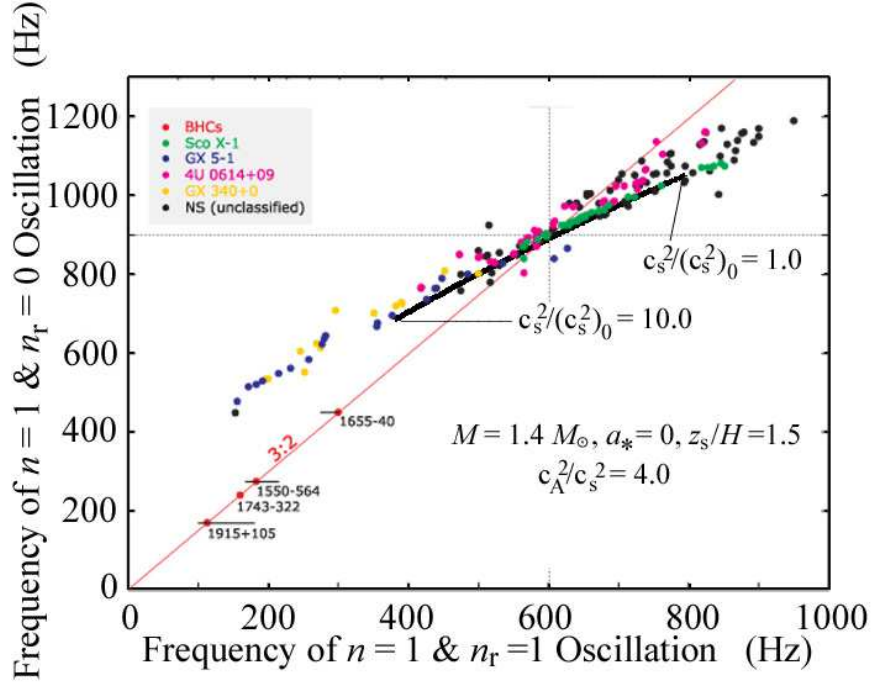

\begin{center}
    \FigureFile(120mm,120mm){figure-5.eps}
\end{center}
\caption{
The same as figure 3, except that the disk is vertically truncated  with $\eta_{\rm s}=1.5$. 
(Color Online)
}
\end{figure}

Here, we consider the case where the disks are terminated in the vertical direction.
To demonstrate rather extreme cases, we adopt $\eta_{\rm s}=1.5$.
The correlation curve is calculated by changing the value of $c_{\rm s}^2/(c_{\rm s}^2)_0$
from 1 to 10, fixing the value of $c_{\rm A}^2/c_{\rm s}^2$.
The case where $c_{\rm A}^2/c_{\rm s}^2=4.0$ is shown in figure 5,
and the case of $c_{\rm A}^2/c_{\rm s}^2=16.0$ is in figure 6.
The results show that decrease of the disk thickness brings about no essential 
characteristic changes to the correlation curves. 

\begin{figure}
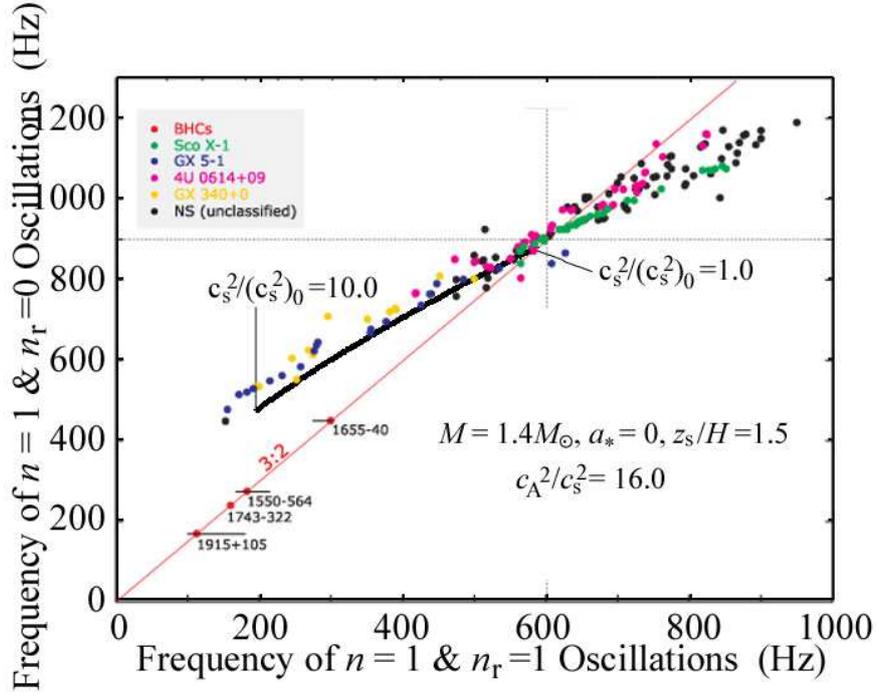

\begin{center}
    \FigureFile(120mm,120mm){figure-6.eps}
\end{center}
\caption{
The same as figure 5 except for $c_{\rm A}^2/c_{\rm s}^2=16.0$.
(Color Online)
}
\end{figure}
\begin{figure}
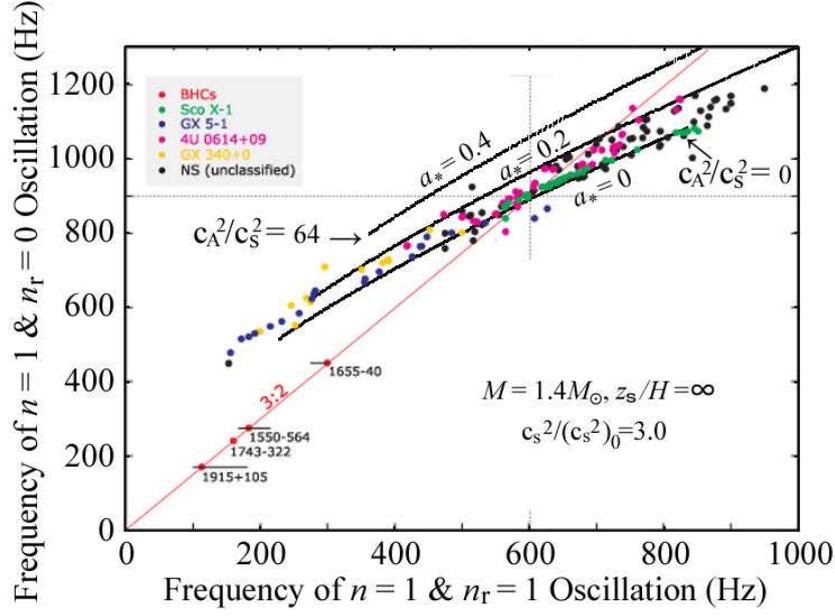

\begin{center}
    \FigureFile(120mm,120mm){figure-7.eps}
\end{center}
\caption{
The same as figure 1 except that cases where spin parameter $a_*$ is non-zero are considered.
Among three correlation curves, the upper one is for $a_*=0.4$, the middle one is for $a_*=0.2$.
The lower one is for $a_*=0$ and the same as that of figure 1.
(Color Online)
}
\end{figure}
\begin{figure}
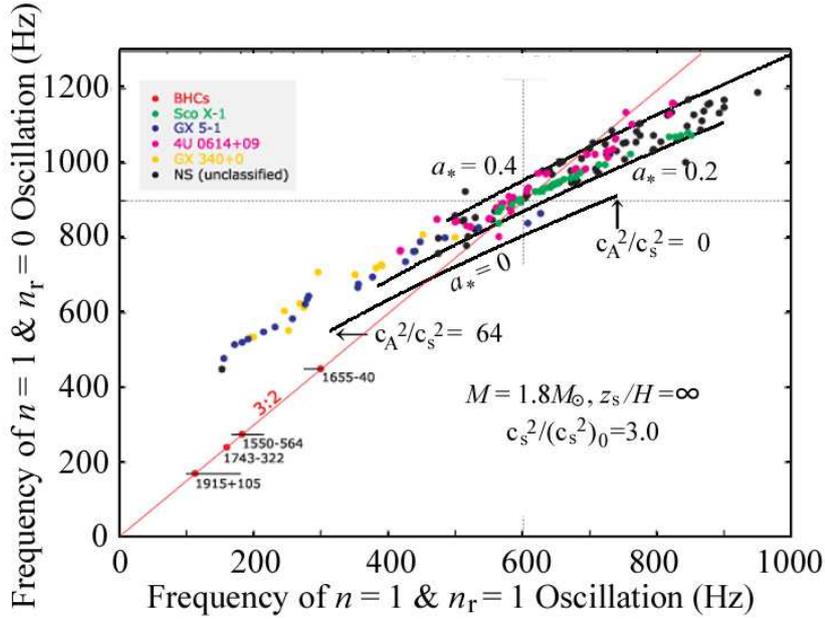

\begin{center}
    \FigureFile(120mm,120mm){figure-8.eps}
\end{center}
\caption{
The same as figure 7 except that the mass of the central source is $1.8M_\odot$.
(Color Online)
}
\end{figure}

Next, we show effects of mass and spin of the central source on the correlation curve (see also paper I).
Figure 7 is for three cases of $a_*=0$, 0.2, and 0.4 with $M=1.4M_\odot$.
Non-truncated disks with $\beta=3.0$ is adopted and $c_{\rm A}^2/c_{\rm s}^2$ is changed from 
$c_{\rm A}^2/c_{\rm s}^2=0$ to 64.
Figure 8 is the same as figure 7 except that $M=1.8M_\odot$ is adopted.\footnote{
Figures 7 and 8 are close to figure 3 in paper I, but the parameter values considered are a little
different.
}
As is easily understood, an increase of spin of the central source increases the frequencies of both the
$n_r=0$ and $n_r=1$ oscillations, but the increase in the $n_r=0$ oscillation is larger compared with
that of the $n_r=1$, since the former is more concentrated in the inner region of disks (figure 8 by Kato
2012).
Hence, the correlation curve shifts in the right-upper direction on the frequency-frequency diagram
as $a_*$ increases, as is shown in figures 7 and 8 (see also figures 3 and 4 in paper I).
An increase of mass of the central source shifts the correlation curve in the left-lower direction on
the diagram, as is also easily understood from the fact that mass increase decreases the frequencies of 
oscillations.
Comparison of figure 7 with figure 8 shows, for example, that if the masses of typical LMXBs presented
in figures are $1.4M_\odot$, their spin will be around $a_*\sim 0$ and at most $a_*<0.2$.
If their masses are $1.8M_\odot$, their spin is larger and $0.2< a_* < 0.4$.

\subsection{Correlation between the $n=1$ (with $n_r=0$) and $n=2$ (with $n_r=0$) Oscillations}

In the previous subsection, we have shown that the correlation curve between the $n_r=0$ and $n_r=1$
oscillaions (both with $n=1$) is almost unchanged on the frequency-frequency diagram for changes of
$\beta$, $c_{\rm A}^2/c_{\rm s}^2$, and $\eta_{\rm s}$, although the range of the curve
depends on parameter ranges.
Different from the above, the correlation between the $n=1$ and $n=2$ oscillations depends strongly
on disk structure.
Here, we demonstrate in details how strong this dependence is.

Let us consider first the case where $\beta$ and $\eta_{\rm s}$ are fixed at some values and
$c_{\rm A}^2/c_{\rm s}^2$ is changed from 0 to 64, which is shown in figure 9.
The mass of the central source is $M=1.4M_\odot$ with no spin, $a_*=0$.
The frequency of the $n=1$ (with $n_r=0$) oscillations is taken on the ordinate, and that of the
$n=2$ (with $n_r=0$) oscillation is on the abscissa.
In this figure the scales of the ordinate and abscissa are changed from those in the previous subsection
in order to superpose the calculated correlation curves on the frequency-frequency diagram 
plotting observational data points (figure 2.9 in a review paper by van der Klis 2004).
The purpose here is to examine whether the $n=2$ (with $n_r=0$) oscillations can describe the observed
HBOs in Z-sources.\footnote{
HBO harmonic and Sub HBO are not of our concern here, since their amplitudes are small and
they will be subsidiary.
}

Except for a nearly vertical curve, five curves are shown in figure 9.
Let us now count the order of the curves, from upper to lower, by the order of the left-end points of
these curves (the third and fourth curves are close each other).
The uppermost one (red curve) is for $\eta_{\rm s}=\infty$ and
$\beta=3.0$, drawn by changing $c_{\rm A}^2/c_{\rm s}^2$ from 4.0 to 64.0.
The right-upper end of the curve is for $c_{\rm A}^2/c_{\rm s}^2=4.0$ and the left-lower
end is for $c_{\rm A}^2/c_{\rm s}^2=64.0$.
The set of ($\eta_{\rm s}$, $\beta$) for the fourth(green) and fifth (blue) curves
are, respectively, (2.0, 3.0) and (1.8, 3.0).
The set of ($\eta_{\rm s}$, $\beta$) for the nearly vertical (magenta) curve on the right side are 
(1.5, 1.0).
[The second (cyan) and third (sienna) curves are mentioned later.]

The above-mentioned curves show that in the non-truncated disk ($\eta_{\rm s}=\infty$) the correlation curve 
runs above the observed plots on the frequency-frequency diagram.
As disks become thin in the vertical direction, the correlation curve moves downward
on the diagram.
However, the gradients of the curves become sharper as $\eta_{\rm s}$ decreases,
and in an extreme case of $\eta_{\rm s}=1.5$, the curve (magenta one) is almost
vertical.\footnote{
In the nearly vertical curve, the upper-right end shifts rightward compared with
other curves.
This is because $\beta=1.0$ is adopted, different from other curves where
$\beta=3.0$ is adopted.
}
That is, these curves cannot well describe observations.

One of possible reasons of this descrepancy between calculated curves and
observations might be that in the above calculations, $c_{\rm A}^2/c_{\rm s}^2$ 
has been taken to be constant independent of $r$.
In real disks the magnetic fields may decrease outwards.
If so, in considering the frequency of the $n=2$ oscillations
this should be taken into account, since the trapped region of the oscillations
is rather wide.
(In trapped $n=1$ oscillations, however, the trapped region is narrow and the effects of radial decrease of
$c_{\rm A}^2/c_{\rm s}^2$ are minor.)
Considering this situation, we have calculated, as a tentative attempt,
the frequencies of the $n=2$ oscillations by assuming that the value of
$c_{\rm A}^2/c_{\rm s}^2$ is constant till $r=4r_{\rm g}$, 
and outside of $4r_{\rm g}$, it decreases as $c_{\rm A}^2/c_{\rm s}^2=(c_{\rm A}^2/c_{\rm s}^2)_0
(4r_{\rm g}/r)^2$, where $(c_{\rm A}^2/c_{\rm s}^2)_0$ is the constant value of 
$c_{\rm A}^2/c_{\rm s}^2$ till the radius of $4r_{\rm g}$.
The correlation  curves obtained by using this $c_{\rm A}^2/c_{\rm s}^2$ are shown
as the second (cyan) and third (sienna) curves for ($\eta_{\rm s}$, $\beta$) being
(2.0, 3.0) and (1.8, 3.0), respectively.
The case of $\eta_{\rm s}=1.8$ and $\beta=3.0$ with the above modified $c_{\rm A}^2/c_{\rm s}^2$
better describe observations compared with other cases, although it is still not satisfactory.

Next, let us consider the case where $c_{\rm A}^2/c_{\rm s}^2$ and $\eta_{\rm s}$ are fixed at some
values and $\beta$ is changed from $\beta=1.0$ to $\beta=10$.
The mass of the central source is again $M=1.4M_\odot$ with no spin, $a_*=0$.
Among six curves in figure 10, the uppermost (red) one is for the non-truncated disk ($\eta_{\rm s}=\infty$)
with $c_{\rm A}^2/c_{\rm s}^2=4.0$.
The curve is above the observed sequence of HBOs on the diagram.
As in figure 9, the curve moves downward as the disk becomes thin in the
vertical direction.
That is, the set of parameters ($\eta_{\rm s}$, $c_{\rm A}^2/c_{\rm s}^2$) in the second
(blue), third (green), fifth (magenta), and sixth (cyan) curves are
(1.8, 4.0), (2.0, 16.0), (1.8, 16.0), (1.5, 4.0).
[Concerning the fourth (sienna) curve we discuss in the next paragraph.]
It is noted that in two cases of the same $\eta_{\rm s}$, the correlation curve calculated with a
larger $c_{\rm A}^2/c_{\rm s}^2$ runs below the curve calculated with a smaller $c_{\rm A}^2/c_{\rm s}^2$.

\begin{figure}
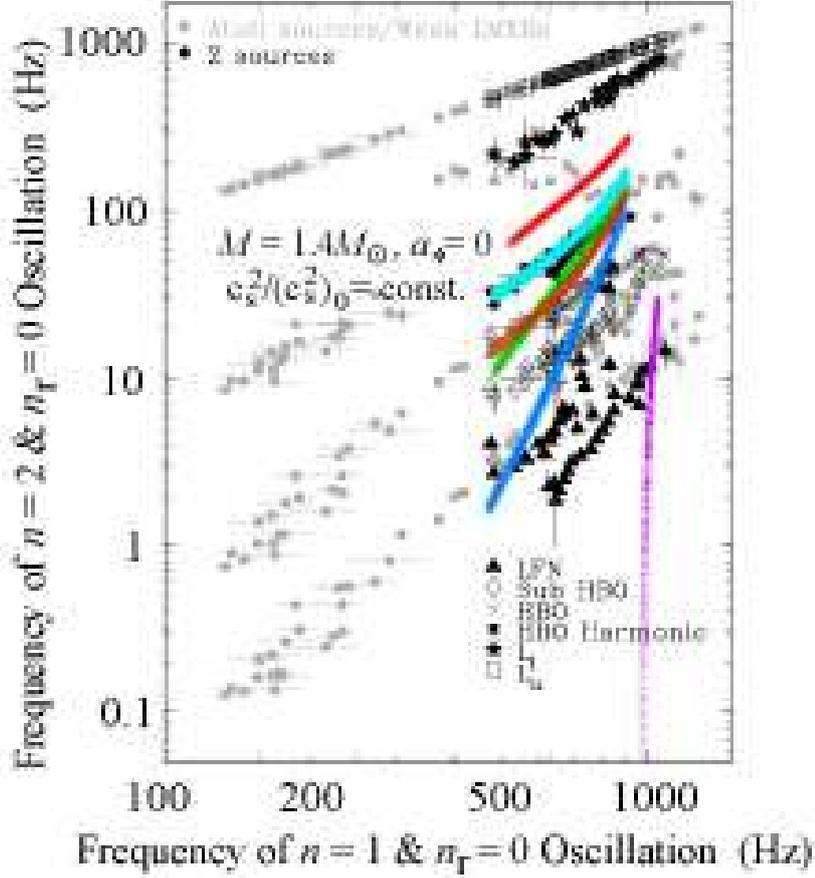

\begin{center}
    \FigureFile(120mm,120mm){figure-9.eps}
\end{center}
\caption{
Diagram comparing the calculated frequency-frequency correlation curve [the frequency of $n=1$ (with $n_r=0$) 
oscillation versus that of $n=2$ (with $n_r=0$) oscillation] with the observed correlation 
between the upper kHz QPO and HBO of Z-sources.
The abscissa is the frequency of $n=1$ (with $n_r=0$) oscillations, and the ordinate is that of
$n=2$ (with  $n_r=0$) oscillations.
The plots of observational data are a part of figure 2.9 by van der Klis (2004).
The cases of $M=1.4M_\odot$ and $a_*=0$ are shown.
The correlation curves are obtained by changing $c_{\rm A}^2/c_{\rm s}^2$ from 
4 to 64 for some cases of $\eta_{\rm s}$ and $\beta\equiv c_{\rm s}^2/(c_{\rm s}^2)_0$.
Let us count the order of curves, from upper to lower, by the order of their left-end points.
In the first (red), fourth (green), and fifth (blue) curves, the set of parameters 
($\eta_{\rm s}$, $\beta$) adopted are 
($\infty$, 3.0), (2.0, 3.0), and (1.8, 3.0).
In the second (cyan) curve, the parameters adopted are the same as those in the fourth (green) one, 
except that  
$c_{\rm A}^2/c_{\rm s}^2$ is taken to be the same constant as in the fourth one till $4r_{\rm g}$, but 
decreases from there outward as $\propto (4r_{\rm g}/r)^2$.
Similarly, in the third (sienna) curve the parameters adopted are the same as those in the fifth (blue) curve,
except that $c_{\rm A}^2/c_{\rm s}^2$ is taken to decrease from $4r_{\rm g}$ outward in the radial direction 
as $\propto (4r_{\rm g}/r)^2$.
The (magenta) curve  which is roughly vertical near the right end of the figure shows the correlation curve in the case where
the set of ($\eta_{\rm s}$, $\beta$) is (1.5, 1.0).
}
\end{figure}
\begin{figure}
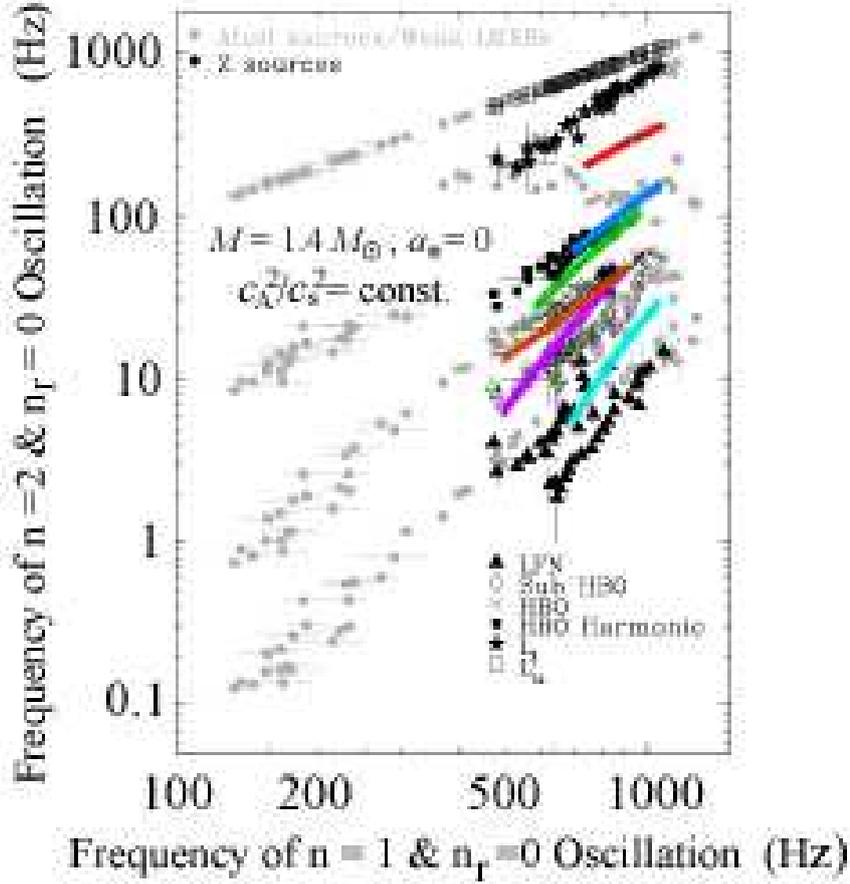

\begin{center}
    \FigureFile(120mm,120mm){figure-10.eps}
\end{center}
\caption{
Similar to figure 9 except that the correlation curves are derived by changing 
$c_{\rm s}^2/(c_{\rm s}^2)_0$ (not $c_{\rm A}^2/c_{\rm s}^2$) from 1.0 to 10.0, 
and $\eta_{\rm s}$ and $c_{\rm A}^2/c_{\rm s}^2$
are fixed at some values as parameters.
Except for the fourth (sienna) curve, 
parameter values of ($\eta_{\rm s}$, $c_{\rm A}^2/c_{\rm s}^2$)
are, from top, ($\infty$, 4.0)(red curve), (1.8, 4.0)(blue curve), 
(2.0, 16.0)(green curve), (1.8, 16.0)(magenta curve), and (1.5,4.0)(cyan curve).
In the fourth (sienna) curve, $c_{\rm A}^2/c_{\rm s}^2$ is not a constant 
in the radial direction.
That is, the value of $c_{\rm A}^2/c_{\rm s}^2$ is 16.0 in the inner region of disk 
as in the fifth (magenta) curve, but is taken to decrease outwards from $4r_{\rm g}$ as 
$c_{\rm A}^2/c_{\rm s}^2\propto (4r_{\rm g}/r)$. 
In all curves, the right-end point of the correlation curve is for $c_{\rm s}^2/(c_{\rm s}^2)_0=1.0$ and 
the left-end point is for $c_{\rm s}^2/(c_{\rm s}^2)_0=10.0$.    
}
\end{figure}

The correlation curves mentioned above cannot well describe observations,
since the correlation curves have larger gradients on the diagram compared with that of observations,
as in cases of figure 9.
Here, however, we should remember again as in the case of figure 9 that for the $n=2$ oscillations 
the trapped region is wide and thus a radial change of $c_{\rm A}^2/c_{\rm s}^2$ should be 
taken into account in calculating their frequencies.
Here, we consider a case where $c_{\rm A}^2/c_{\rm s}^2$ decreases outside of $4r_{\rm g}$ as
$c_{\rm A}^2/c_{\rm s}^2=(c_{\rm A}^2/c_{\rm s}^2)_0(4r_{\rm g}/r)$, where $(c_{\rm A}^2/c_{\rm s}^2)_0$
is the constant value of $c_{\rm A}^2/c_{\rm s}^2$ inside of $r=4r_{\rm g}$.
The correlation curves in the case of $\eta_{\rm s}=1.8$ and $(c_{\rm A}^2/c_{\rm s}^2)_0=16.0$
are shown as the fourth (sienna) curve.
This case seems to better describe observations.

\section{Summary and Discussion}

In this paper we have examined the possibility that the two-armed ($m=2$) vertical p-mode oscillations
trapped in the innermost region of magnetized disks are the origin of 
kHz QPOs and HBO in LMXBs.
The disks are assumed to be isothermal in the vertical direction but truncated at a certain height.
The magnetic fields are assumed to be toroidal.
More concretely speaking, we suggested that i) two oscillations which are both fundamental in
the vertical direction (i.e., $n=1$) but the fundamental ($n_r=0$) and first overtone ($n_r=1$)
in the radial direction are twin kHz QPOs and 
ii) the oscillation which is the first overtone in the vertical direction  (i.e., $n=2$)
but the fundamental in the radial direction (i.e., $n_r=0$) is HBO.
To examine this possibility we have compared i) the calculated frequency correlation between the
$n_r=0$ and $n_r=1$ oscillations (both with $n=1$) with the plots of observed twin kHz
QPOs on frequency-frequency diagram, and ii) the calculated frequency correlation between the $n=1$
and $n=2$ oscillations (both with $n_r=0$) with the observed correlation between kHz QPOs and
HBO on frquency-frequency diagram.

We found that the frequency correlation between the $n_r=0$ and $n_r=1$ oscillations (both with $n=1$) can
well describe the observed correlation of twin kHz QPOs.
That is, in the present disk model, there are three parameters specifying disk structure,
i.e., $\beta\equiv[c_{\rm s}^2/(c_{\rm s}^2)_0]$, $c_{\rm A}^2/c_{\rm s}^2$, and $\eta_{\rm s}
\equiv z_{\rm s}/H$.
Changes of these disk parameters change the frequencies of both oscillations of $n_r=0$ and $n_r=1$
(with $n=1$), and we have a correlation curve on frequency-frequency diagram.
We find that the correlation curve is almost along an
unique curve, independent of disk parameters (see figures 1 to 6), as long as $M$ and $a_*$
are fixed.
It is noted, however, that the range of correlation curve depends on range of changes 
of disk parameters (see figures 1 to 6).

In Z-sources of LMXBs, HBOs are often observed and their time variations are correlated with those
of kHz QPOs (Psaltis et al. 1999).
In this paper, we have examined whether the observed correlation between kHz QPOs and HBOs can be
described in the framework of two-armed vertical p-mode oscillations.
We adopt the picture that the oscillation which is the first overtone in the vertical direction 
(i.e., $n=2$) and the fundamental in the radial direction (i.e., $n_r=0$) is HBO.
Observations show that the amplitude of HBOs is much larger than those of kHz QPOs.
This will be accounted for in this model, since 
as was shown before (Kato 2011a, 2012), the $n=2$ oscillations have  
wide propagation regions.

From figures 9 and 10, we see that to describe the frequency correlation between kHz QPOs and HBOs 
by our present model,
the disks on HB of Z-sources must be at least truncated in the vertical thickness.
This suggests that the disks on HB must be surrounded by hot coronae.
Figures 9 and 10 further suggest that if HB is a sequence of changes of 
magnetic fields or disk temperature, $c_{\rm A}^2/c_{\rm s}^2$ in disks  
must decrease outwards.
Furthermore, the picture of HB being a sequence of time change of disk temperature will
be better than that of time change of disk magnetic fields, since the curves in figure 10 
seem to be closer to the observed ones compared with those in figure 9.
To quantitatively examine whether our model can describe observed HBOs, 
however, more realistic disk
models and detailed numerical calculations will be necessary.
This is because in our model many assumptions and simplifications are involved.
For example, the disks are assumed to be vertically isothermal and $c_{\rm A}^2$ is also constant
in the vertical direction.
The former will be allowed as the first approximation, since the disks which are terminated 
in the vertical direction by hot corona will be close to isothermal.
It is, however, uncertain how much the assumption of $c_{\rm A}^2=$ const. in the vertical 
direction is realistic and how the final results are modified when $c_{\rm A}^2/c_{\rm s}^2$ 
changes in the vertical direction.
Radial dependences of $c_{\rm A}^2/c_{\rm s}^2$, $c_{\rm s}^2/(c_{\rm s}^2)_0$, and $\eta_{\rm s}$
also should be considered, especially when the $n=2$ oscillations are examined.  

Z-sources are known as the brightest group of NS LMXBs which have large accretion rate close 
to or above the Eddington limit.
In early studies, the mass accretion rate is supposed to increase monotonically in the direction
HB-NB-FB (e.g., Priedhorsky et al. 1986).
Recent spectral studies by the extended ADC (accretion disk corona) model on GX 340+0 (Church et al. 2006)
and GX 5-1 (Jackson et al. 2009), however, suggest the opposite trend (Church et al. 2008).
That is, in Church et al.'s model based on their spectral analyses, 
the soft apex between NB and FB is a state in which the mass accretion rate 
$\dot {m}$ is low, and $\dot {m}$ increases on NB to HB.
In their model, due to strong radiation from central NS, which results from  large $\dot{m}$,
the innermost part of accretion disks on the upper NB and HB are supposed to be desrupted.
They further supopose that this desruption is the causes of decrease of the frequency of 
kHz QPOs on HB aparting from the 
hard apex and of appearance of radio emission in the upper NB and HB (Penninx 1989).

The disk structure of HB suggested by our present results is similar to that suggested by
Church et al. (2008) in the sense that optically thick disks on HB must have high temperature and be
geometrically thin by horizontal truncation (perhaps by the presence of corona).
Concerning the direction of increase of mass accretion rate on HB, our results are insufficient to say
anything definitively, but suggest an increase of mass accretion rate along HB
apart from hard apex, since as mentioned before, in our results HB would be a sequence of 
disk temperature increase apart from hard apex.@ 
It will be of importance to notice here that geometrically thin, high temperature 
disks with strong magnetic fields are realized, as disks bridging between ADAFs and optically thick 
standard disks (or slim disks), when mass accretion rate is close to or above
the Eddington limit (Machida et al. 2004; Oda et al. 2007, 2009, 2010).
 
It is noted that we suppose there is no appreciate outward retreat of the inner edge of
geometrically thin disks during the evolution along HB.
We have calculated frequency changes of $n_r=0$ and $n_r=1$ (both with $n=1$) oscillations
by changing the radius of the inner edge, although the results are not shown in the text.
The results show that their correlation curve on the frequency-frequency diagram has a sharper  
gradient (close to 3:2) than that of the observed twin kHz QPOs. 

Excitation of the trapped disk oscillations is a problem remained to be examined.
The most conceivable process is stochastic excitation by turbulence (Goldreich \& Keely 1977a, b).
This process is now known as the main cause of solar and stellar non-radial oscillations.
Similar processes will be expected in disks, where much stronger turbulence is expected
compared with inside the stellar convection zone.

\bigskip
\leftskip=20pt
\parindent=-20pt
\par
{\bf References}
\par
Aliev, A.N., \& Galtsov, D.V. 1981, General Relativity and Gravitation, 13, 899\par
Church, M.J., Halai, G.S., \& Ba\l ci\'{n}ska-Church, M. 2006, A\&A, 460, 233 \par
Church, M.J., Ba\l ci\'{n}ska-Church, M. Jackson, N.K., \& Gibiec, A. 2008, 
    in VII Microquasar Workshop: Microquasars and Beyond (at Foca, Izmir, Turkey)
    (arXiv:0811.2680v1)\par
Goldreich, P, \& Keely, D.A. 1977a, ApJ, 211, 934\par 
Goldreich, P, \& Keely, D.A. 1977b, ApJ, 212, 243\par 
Jackson, N.K., Church, M.J. \& Ba\l ci\'{n}ska-Church, M. 2009, A\&A, 494, 1059  \par
Kato, S. 2001, PASJ, 53, 1\par 
Kato, S. 2010, PASJ, 62, 635 \par
Kato, S. 2011a, PASJ, 63, 125 \par
Kato, S. 2011b, PASJ, 63, 861 (paper I) \par
Kato, S. 2012, PASJ, to be published \par
Kato, S., Fukue, J., \& Mineshige, S. 1998, Black-Hole Accretion Disks 
  (Kyoto: Kyoto University Press), chap. 17 \par
Kato, S., Fukue, J., \& Mineshige, S. 2008, Black-Hole Accretion Disks --- Towards a New paradigm --- 
  (Kyoto: Kyoto University Press), chaps. 3 and 11 \par
Machida, M., Nakamura, K.E., \& Matsumoto, R., 2004, PASJ, 56, 671 \par
McClintok, J.E., Narayan, R., Davis, S.W., Gou, L., Kulkarni, A., Orosz, J.A.,
   Penna, R.F., Remillard, R.A., \& Steiner, J.F. 2011, 
   in Classical and Quantum Gravity; Special volume of GR19,
   eds. D.Marolf \& D.Sudarsky (arXiv 1101.0811) \par
Oda, H., Machida, M., Nakamura, K.E., Matsumoto, R. 2007, PASJ 59, 457 \par 
Oda, H., Machida, M., Nakamura, K.E., Matsumoto, R. 2009, ApJ, 697, 16 \par 
Oda, H., Machida, M., Nakamura, K.E., Matsumoto, R. 2010, ApJ, 712, 639 \par 
Okazaki, A.T., Kato, S., \& Fukue, J. 1987, PASJ, 39, 457\par
Penninx, W. 1989, in J.Hunt, \& B.Battrick, eds. "Proceedings of 
    the 23rd ESLAB symposium on Two Topics in X-ray Astronomy", Bologna, Sept. 1989,
    ESA publications ESA SP-296, 185 \par
Priedhorsky, W., Hasinger, G., Lawin, W.H.G., et al. 1986, ApJ, 306,L91 \par
Psaltis, D., Belloni, T., van der Klis, M. 1999, ApJ, 520, 262\par 
van der Klis, M. 2004, in Compact stellar X-ray sources (Cambridge University Press), 
   eds. W.H.G. Lewin and M. van der Klis (astro-ph/0410551)    \par

\bigskip\bigskip
\leftskip=0pt
\parindent=-20pt

\noindent
Note added on Feb. 3, 2012:

\noindent
In this paper we have calculated the frequency -- frequency correlations, based on the
approximation (Kato 2012) that the vertical p-mode oscillations are nearly vertical and
thus the horizontal motions associated with them are small perturbations over the vertical ones.
This approximation is qualitatively relevant, but not accrate enough.
In the special case where there is no magnetic field, we can calculate the frequencies of the
vertical p-mode oscillations without using the approximation.
In this special case we can thus compare the correlation curves based on the approximation with
those without the approximation.
The comparison shows that the correlation curves in the present paper based on the
approximation are almost the same as those without the approximation, although the parameter values
describing a almost same point on correlation curve are different.
This will be presented in a subsequent paper.

\end{document}